# The Effects of Carboxylic Acids on the Aqueous Dispersion and Electrophoretic Deposition of $ZrO_2$

*Dorian A. H. Hanaor,[a,*] Marco Michelazzi[b], Cristina Leonelli[b], Charles C. Sorrell[a]*

[a]*School of Materials Science and Engineering, University of New South Wales, Sydney, NSW 2052, Australia*

[b]*Department of Materials and Environmental Engineering, University of Modena and Reggio Emilia, 41100 Modena, Italy*

*\* E-Mail: dorian@ unsw.edu.au*

## Abstract

The agglomeration, electrokinetic properties and electrophoretic deposition behaviour of aqueous suspensions of $ZrO_2$ with carboxylic acid additives were studied in comparison with conventional pH adjustment. It was found that citric acid imparted negative zeta-potential values and electrosteric stabilisation to particles in suspensions at all pH levels. The examination of additions of carboxylic acids to $ZrO_2$ suspensions revealed that these reagents cause a sharp drop in zeta-potential at distinct addition levels, which correspond to surface saturation of the particles with negatively charged carboxylate groups. Adsorption cross sections of citric acid, EDTA and oxalic acid were evaluated from these results, showing that both citric acid and EDTA coordinate to $ZrO_2$ surfaces by two carboxylate groups while oxalic acid is coordinated by one group. The use of carboxylic acids was shown to facilitate superior electrophoretic deposition in comparison with zeta-potential modification by conventional pH adjustment through improved suspension stability.

## 1. Introduction

The electrokinetic properties of oxide particles in aqueous suspensions are of paramount importance in controlling the electrophoretic processes involving these materials. Such processes include the electrophoretic deposition (EPD) of thick ceramic films, [1-5] the filtration and separation of oxide particles, [6,7] and the removal of solid contaminants from soil, [8,9] Further, electrokinetic properties are important parameters in governing rheological properties of thick suspensions or slurries used for slip casting, screen printing, gel casting, direct coagulation casting (DCC), and extrusion. [10-13]

EPD in particular is of growing importance owing to the capacity of this method to be used to fabricate unique microstructures in a variety of forms from dilute suspensions of fine particles in a cost-effective manner. [1, 3, 14-16] EPD consists of the movement of charged particles under an electric field and their compact deposition on a substrate. Typically, this deposition is followed by a densification step through heat treatment. EPD often is conducted using organic suspension media owing to the occurrence of the parasitic process of water electrolysis in EPD from aqueous suspensions. Despite the associated problem of water electrolysis, the use of aqueous suspension media for EPD is attractive due to the lower environmental impact, greater simplicity, and lower cost that facilitate the application of such methods in larger scale processes.

The electrokinetic properties of a particle in suspension are governed by the electric charge distribution in the double layer that surrounds the particle. [17, 18] This double layer is formed when a surface-charge-carrying solid particle suspended in a liquid becomes surrounded by counter-ions of charge opposite to that of the particle surface. As the particle moves in the solution, the plane beyond which counter-ions do not migrate along with the particle is known as the slipping plane. The electrical potential at the slipping plane is known as the zeta potential ($\zeta$) and typically is measured in mV.

As a result of increased electrostatic repulsion inhibiting agglomeration and settling, a suspension of particles showing a high absolute value of zeta potential is more stable in comparison to suspensions exhibiting lower zeta potential absolute values. In electrophoretic processes, a high zeta potential is desirable as it enhances the rate of particle movement under a given electrical field while inhibiting the sedimentation of the material. The point at which the potential at the slipping plane is zero is known as the isoelectrc point (IEP). At this point electrostatic repulsion is minimised and Van der Waals forces facilitate agglomeration.

The rate of particle movement under an electrical field is known as the electrophoretic mobility, µ. This parameter is defined by equation 1. [1, 19]

$$\mu = v/E \qquad (1)$$

Here, v = velocity and E = electric field. Electrophoretic mobility is the key parameter in governing the kinetics of EPD. Particle movement under an electric field is directly related to the magnitude of the zeta potential. This can be seen in the expression of electrophoretic shown in equation 2. [1, 15]

$$\mu = \frac{\varepsilon_0 \varepsilon_r \zeta}{\eta} \qquad (2)$$

Here, $\varepsilon_0$ is the permittivity of free space, $\varepsilon_r$ and $\eta$ are the permittivity and viscosity of the suspension medium, respectively, and $\zeta$ is the zeta potential of the suspended particles. This equation is an approximation and it assumes that the double layer thickness is negligible relative to the particle diameter. It can be seen that, for a given suspension medium, the electrophoretic mobility is proportional to the zeta potential.

For given deposition conditions, the kinetics of EPD in planar geometries are governed by the electrophoretic mobility (or zeta potential) of the suspended particles as shown by the Hamaker equation in equation 3. [2, 20]

$$d = C_S \mu E t \qquad (3)$$

Here d is the density of the deposit (g cm$^{-2}$), $C_s$ is the solids loading (g cm$^{-3}$), µ is the electrophoretic mobility (cm$^2$ s$^{-1}$ V$^{-1}$) (which can be expressed as a function of ζ as shown in Eq. 2.), E is the electric field (V cm$^{-1}$) and t is time (s).

Zeta potential is varied most commonly by pH adjustment. The zeta potential generally goes to more positive values with decreasing pH level. [1, 21] The point at which ζ=0, the IEP of a particulate suspension, generally is discussed in terms of the pH at which this occurs. It should be noted that the IEP of a particular powder can occur at different pH levels through the use of dispersants or by the use of alternative pH adjustment agents. [22-24]

Organic additives often are used as dispersants to increase repulsive forces between ceramic particles in suspension. Typically, these additives are charge-carrying long-chain polyelectrolytes. These additives adsorb on particle surfaces and modify the surface charge of these particles in suspension and thus enhance the interparticle electrostatic repulsion while further providing a steric barrier to agglomeration. Commonly used polyelectrolyte dispersants typically have molecular weights in the range 6,000-15,000. [3, 10, 25-28] Dispersants of lower molecular weight may have advantages over such polyelectrolytes owing to their low costs, higher adsorption capacities, lesser effects on post-firing microstructure and lower environmental impact both in polar and non polar solvents. [11, 23, 24, 29] Carboxylic acids have been shown to act as low molecular weight dispersants for aqueous suspensions of alumina. [11, 22, 24] Such reagents are reported to impart negative surface charge on particles in suspension through the surface adsorption of the carboxylate anion (RCOO$^-$). Carboxylic acids have been used to facilitate anodic EPD with decreased levels of water electrolysis in acidic aqueous suspensions of $TiO_2$. [20]

$ZrO_2$ is an oxide of considerable technical interest owing to its potential applications in structural ceramics, bioceramics, oxygen sensing materials and in electrolyte films in solid oxide fuel cells (SOFCs). EPD of this oxide has been shown to be one of the most promising fabrication methods in the production of SOFCs. [18, 30-33] Studies of the zeta potential behaviour of undoped $ZrO_2$ have shown the IEP of this material to be at an pH level of ~5.5 [15, 34] while tetragonal zirconia doped with 3% yttria (as used for SOFCs) has been reported to exhibit an IEP at pH~7. [35] The present work discusses the electrokinetic properties of aqueous suspensions of monoclinic $ZrO_2$ dispersed with the aid of carboxylic acids and the consequent effects of such dispersion methods on the electrophoretic deposition behaviour of this material.

## 2. Materials and Methods

Commercial monoclinic zirconia powder (Colorobbia, Italy) was used for suspension preparation. The powder was hand-ground using an agate mortar and pestle for degglomeration. Surface area of the degglomorated powder was determined using $N_2$ adsorption isotherms at 77k in conjunction with BET calculation methods. A 0.1 wt% (0.001 g/mL) suspension of $ZrO_2$ in distilled water was stirred magnetically and sonicated in an ultrasonic bath for 10 minutes to achieve consistent dispersion of particles in suspension.

Zeta potential, electrophoretic mobility, suspension conductivity and agglomerate size distribution were measured using a Nano-Zetasizer (Malvern Instruments, Worcestershire, UK). Reagent-grade anhydrous citric acid (Sigma Aldrich, USA), oxalic acid, and ethylene diamine tetra-acetic acid (EDTA) (both Univar, Germany) were used as carboxylic dispersants while nitric acid (70 %), ammonium hydroxide (25%) and sodium hydroxide (all Univar) were used for conventional pH adjustment.

Zeta potential and electrophoretic mobility variation with conventional pH adjustment of $ZrO_2$ suspensions was compared to data resulting from pH adjustment using citric acid and sodium hydroxide for pH adjustment. At all pH values, citric acid was a component of the solution in order to ensure that the citrate group was available for adsorption.

The dispersant effects of three carboxylic acids with differing numbers of carboxyl groups were investigated through additions of controlled small quantities of dilute citric acid, oxalic acid, and EDTA to suspensions which had at fixed pH levels adjusted by prior addition of nitric acid. This was done in order to examine the dispersion phenomena of $ZrO_2$ suspensions by determining the electrokinetic properties as a function of additive concentration. pH levels of the dilute additive solutions of carboxylic acid were measured as pH=3.2, pH=2.9 and pH=3.4 for Citric acid, oxalic acid and EDTA respectively. Owing to the small molar quantities of carboxylic additives, only a slight variation in ionic strength occurred as supported by the lack of significant variation in suspension conductivity.

The anodic electrophoretic deposition of $ZrO_2$ was facilitated at low pH levels using citric acid and oxalic acids for pH/ ζ modification, and at high pH levels using sodium hydroxide. Cathodic EPD was carried out using nitric acid for pH/ ζ modification. Electrophoretic depositions were carried out from suspensions of 0.01g $cm^{-3}$ solids loadings onto graphite substrates (GrafTech International, Ohio, USA) cut to approximate dimensions 25X25x2 mm and masked to leave one side available for deposition. Depositions were carried out for 10 minutes at 10V with an electrode separation of 20mm. EPD results were evaluated by determining the weight of $ZrO_2$ deposited per unit of area.

## 3. Results

### 3.1. Characterisation of $ZrO_2$ Powder

Adsorption isotherms of $N_2$ at 77K in conjunction with BET analysis methods revealed the deglommorated powder to exhibit a surface area of 121.7 $m^2g^{-1}$. The size distribution of dispersed $ZrO_2$ particles in aqueous suspension as determined by dynamic light scattering is shown in **Fig. 1.**

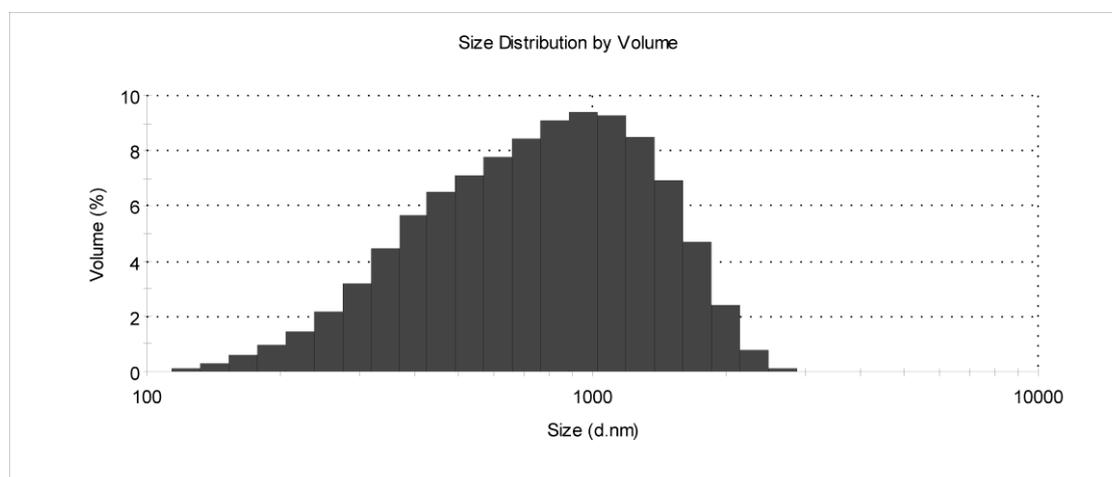

**Fig. 1. Size distribution histogram by volume for dispersed monoclinic $ZrO_2$ powder**

*3.2. Variation of zeta potential with pH*

The conventional variation of zeta potential and a particle dispersion as a function of pH (without effects of surface adsorbed dispersants) in aqueous suspensions of $ZrO_2$ was determined using nitric acid and ammonium hydroxide for pH adjustment. This is shown in **Fig.2.** These regents are unlikely to adsorb on oxide surfaces due to the absence of functional groups which are able to substitute for the surface hydroxo groups.[36] In consistency with Eq. 2, data showed that zeta potential varied as a linear function of electrophoretic mobility, therefore µ values are not reported. The agglomerate sizes shown in the figure are approximate. The IEP occurs at pH ~5.5, which is similar to values reported elsewhere for non-stabilized zirconia [15, 26]. Soft-agglomerate formation of the $ZrO_2$ particles in suspension was observed to occur when the absolute value of the zeta potential was less than ~30 mV, which is observed commonly in many oxide suspensions [37, 38]. As expected, agglomeration was maximised in the region of pH ~5.5, which corresponds to the pH at which the zeta potential is minimal. At high (>9) and low (<4) pH levels a moderate decrease in zeta potential (absolute value) occurs. This is an anticipated consequence of the greater ionic strength imparted by higher reagent concentrations required to adjust the pH values to strongly acidic and basic conditions and the resultant compression of the electric double layer. [39, 40]

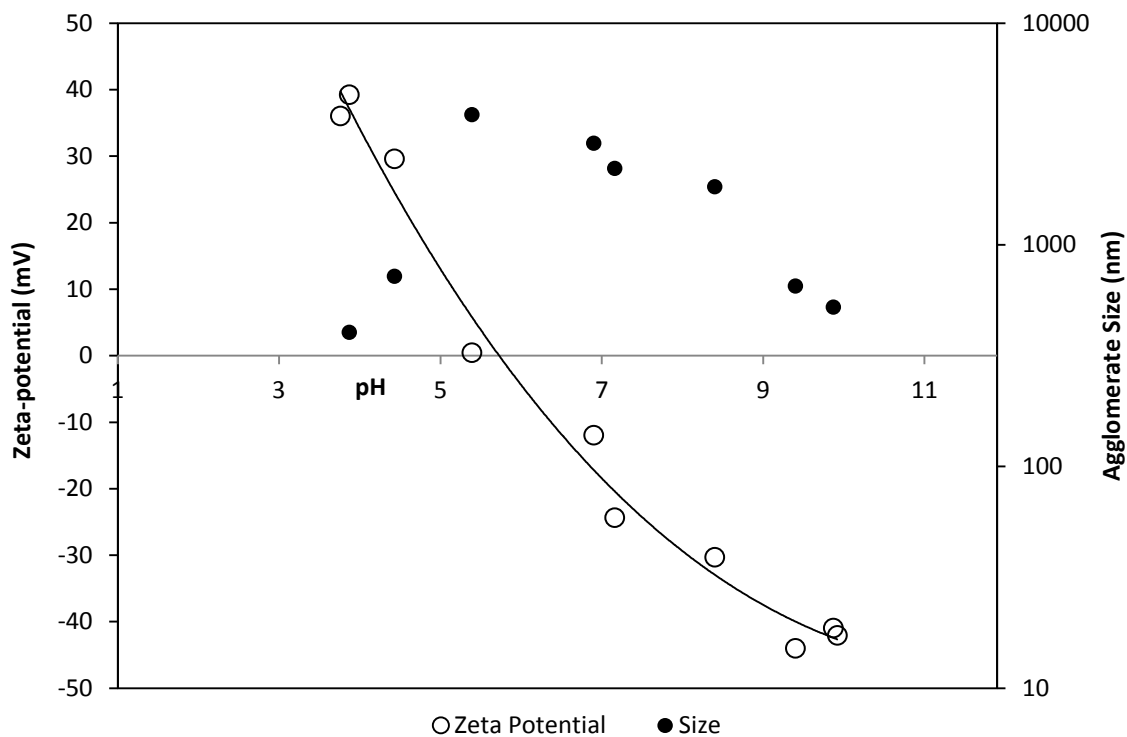

**Fig.2. Conventional zeta potential and agglomeration behaviour of monoclinic $ZrO_2$ as a function of pH.**

In comparison to the zeta potential variation shown in Fig. 2, which is consistent with the variation in zeta potential with pH reported elsewhere, the use of citric acid led to a significantly different zeta potential variation as a function of pH, as shown in **Fig. 3**. It can be seen that the zeta potential remains negative even at low pH values. At all pH levels particle dispersion was maintained and no agglomeration of the suspended particles was observed, suggesting that a zeta potential in the region of only ~18 mV (as limited by the dissociation constant of citric acid $pK_{a1}$) is sufficient to maintain particle dispersion. The negative zeta potential values are consistent with previous observations of the dispersion of oxide suspensions using carboxylic acids. [11, 22, 24, 41]

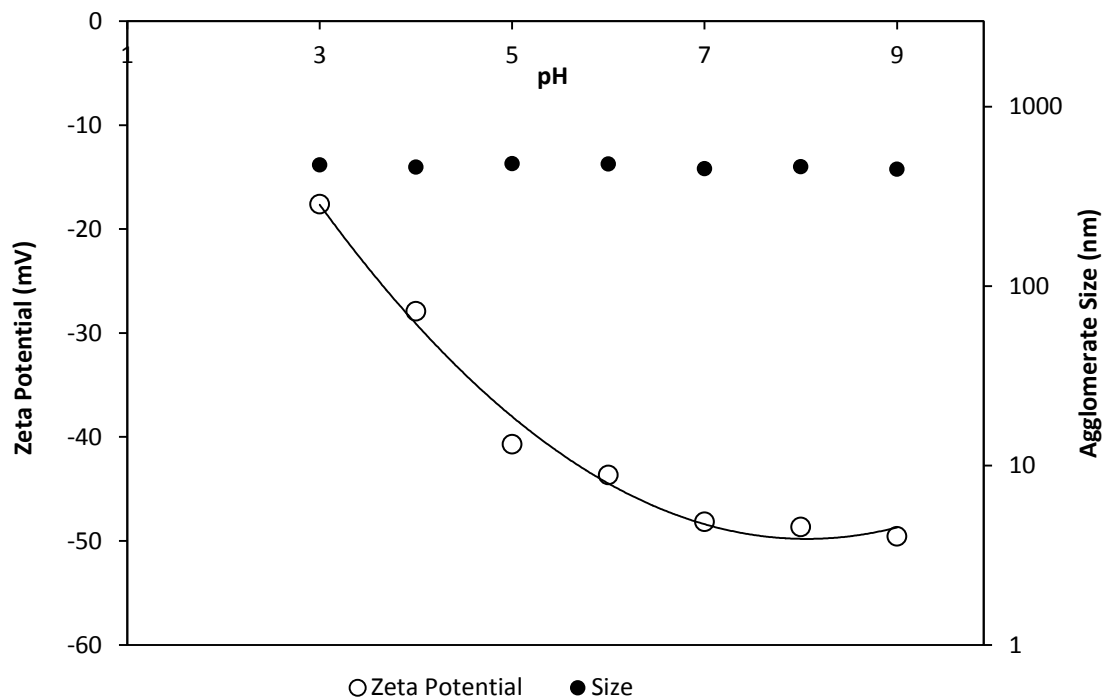

**Fig. 3.** Zeta potential and agglomeration behaviour of $ZrO_2$ as a function of pH varied with citric acid and sodium hydroxide.

*3.3. Effects of carboxylic acids on zeta potential*

The dispersant effects of different carboxylic acids on aqueous suspensions of $ZrO_2$ were investigated by measuring the zeta potential of acidified and native-pH suspensions with controlled additions of the three carboxylic acids oxalic acid, citric acid and EDTA, which exhibit two, three and four carboxyl groups respectively with symmetrical stereochemistry.

Aqueous $ZrO_2$ suspensions were adjusted to an initial acidity of pH=4 with nitric acid (in order to be able to lower an initially positive zeta potential across the IEP) and then treated with additions of dilute solutions of carboxylic acids. As shown in **Fig. 4**, the quantities of carboxylic acids added were sufficient to decrease the zeta potential significantly; this resulted in a pH change of only ±0.2. Repetition of these experiments confirmed these results. It can be seen that a higher concentration of oxalic acid is required to facilitate dispersion in comparison with citric acid or EDTA. Citric and oxalic acids yielded similar and lower ultimate zeta potential values than that for EDTA.

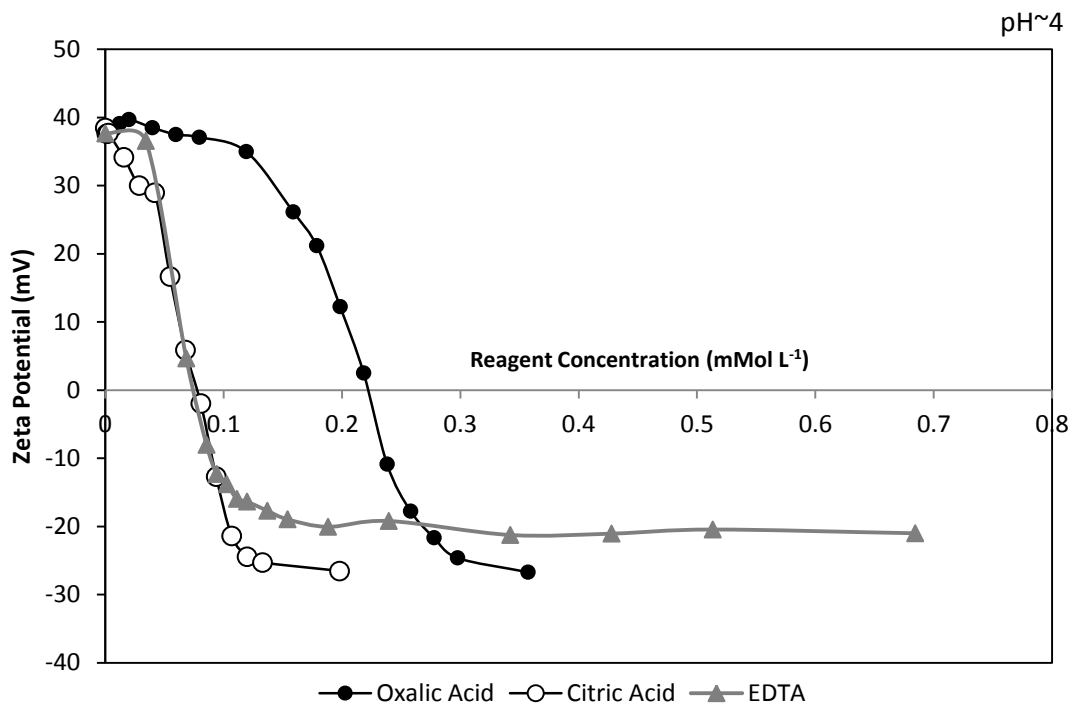

**Fig. 4. The effects of carboxylic acid additions on the zeta potential of $ZrO_2$ at pH =4.**

As shown in **Fig. 5**, similar additions of these carboxylic acids were made without pH standardisation (native-pH). The starting suspension, comprised only of $ZrO_2$ powder in distilled water, was found to have an initial pH of ~5.4, which can be compared to the data point in Fig. 1 at which the zeta potential was close to zero (no addition of acid or base). The acidic native-pH for $ZrO_2$ has been observed elsewhere. [15, 26] The onset of the drop in zeta potential occurred immediately and at a lower concentration compared to the addition levels to $ZrO_2$ suspensions at pH=4. It also can be seen that the curves are not as smooth as those in Fig. 3. It is likely that both of these effects occurred because, as shown in Fig. 2, at pH=4, dispersion is optimised and so the suspension is more resistant to the small changes in pH that occur due to the additions while at the native pH level, dispersion is minimised as this pH value is close to the IEP and so the suspension responds more readily to small changes in pH.

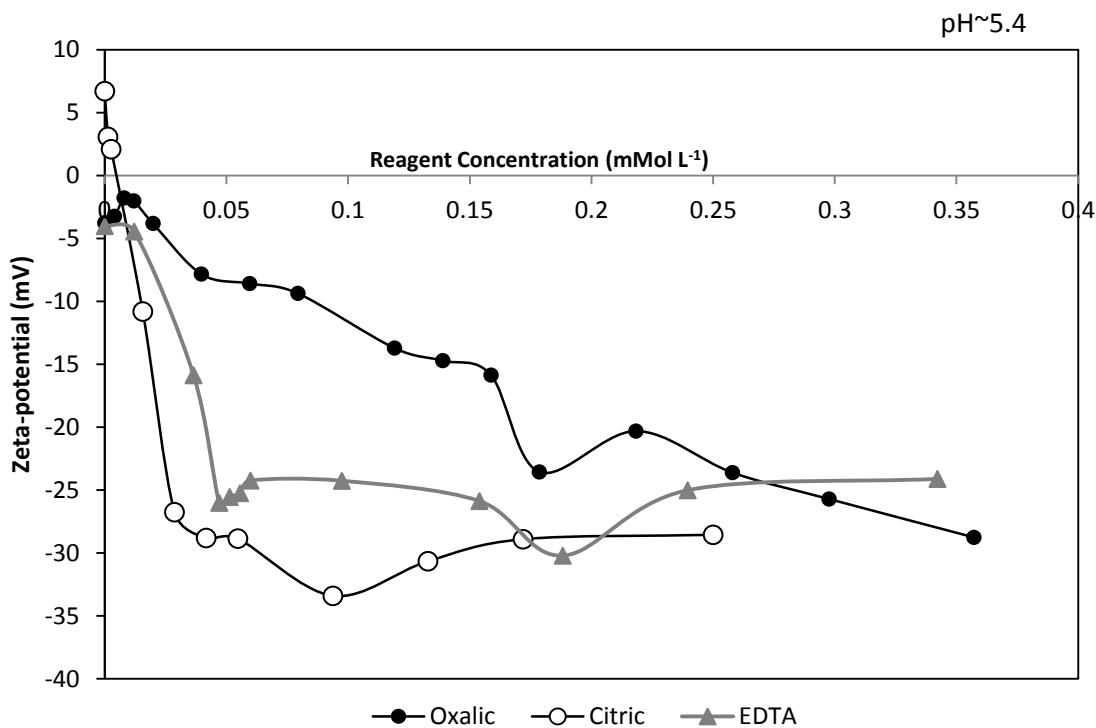

**Fig. 5. The effects of carboxylic acids on the zeta potential of $ZrO_2$ at its native pH level of ~5.4.**

Citrate adsorption on $ZrO_2$ particles was further examined by measuring the variation in pH of a $ZrO_2$ suspension during controlled additions of dilute citric acid. Both the suspension and the citric acid additive were prepared at pH=4 to eliminate acid-base neutralisation effects. It was found that as the reagent was added an increase in pH occurred, reaching a maximum as the zeta potential reached a minimum as shown in **Fig. 6**. These results indicate that the change in pH of the suspension is a consequence of the interaction of citric acid with the suspended $ZrO_2$ particles

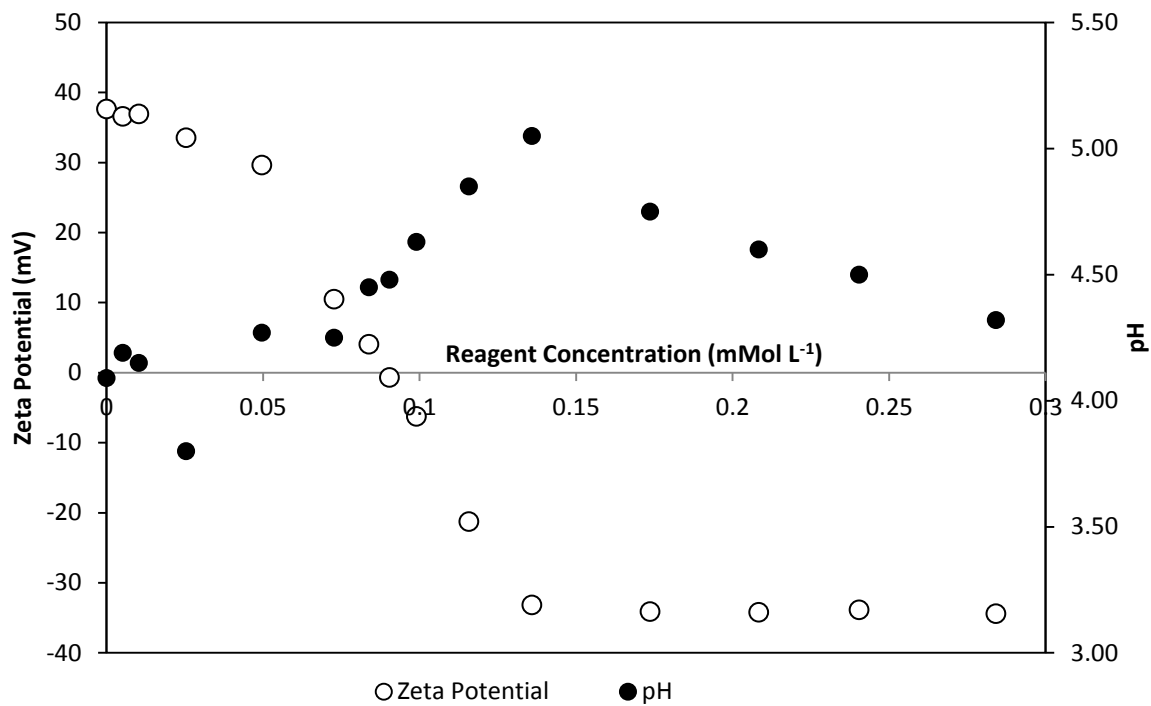

**Fig. 6.** Variation of zeta potential and pH of a $ZrO_2$ suspension (initial pH=4) with additions of citric acid also at pH=4.

*3.4. Electrophoretic deposition*

The effects of carboxylic acids on the EPD of $ZrO_2$ were examined by anodic and cathodic depositions and are outlined in **Table 1**. Citric and oxalic acids improved suspension stability and facilitated effective deposition of thick films on the graphite substrates. In contrast it was found that despite high absolute values of electrophoretic mobility (and zeta potential), using nitric acid and sodium hydroxide for pH adjustment in cathodic and anodic depositions respectively resulted in low deposit densities due to settling of $ZrO_2$ particles in the suspension. In addition to the sedimentation of $ZrO_2$, EPD with nitric acid resulted a frothing of the suspension at the cathode which caused further deterioration to the integrity of the deposit. A comparison of calculated deposit mass per unit area predicted using the Hamaker equation (Eq. 1.) is illustrated in **Fig. 7.**

**Table 1. Electrophoretic deposition data.**

| Reagent | pH | EPD Type | Mobility ($10^{-4} cm^2 V^{-1} S^{-1}$) | Deposit Mass (mg cm$^{-2}$) | Calculated Mass (mg cm$^{-2}$) |
|---|---|---|---|---|---|
| Citric Acid | 3.43 | Anodic | -1.79 | 4.96 | 5.37 |
| Oxalic Acid | 3.57 | Anodic | -2.02 | 4.48 | 6.06 |
| Nitric Acid | 2.98 | Cathodic | 2.83 | 1.03 | 8.49 |
| Sodium Hydroxide | 9.88 | Anodic | -3.35 | 0.91 | 10.05 |

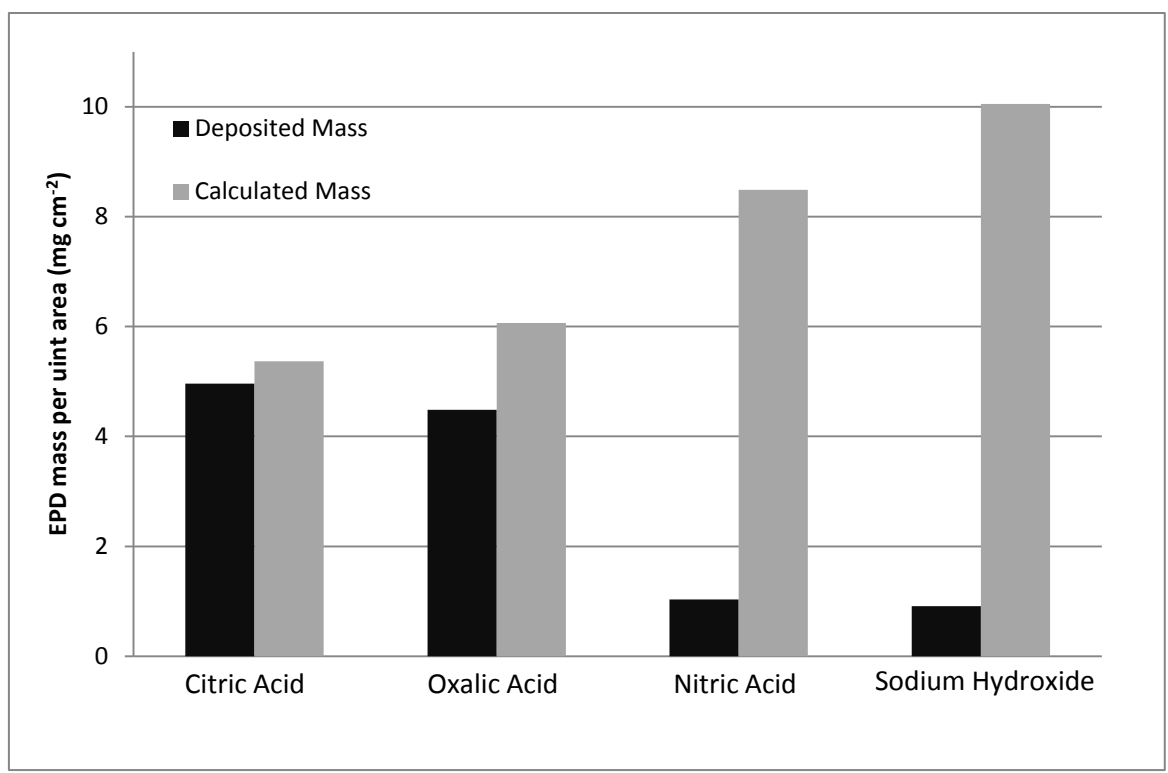

**Fig. 7.** EPD of ZrO$_2$: a comparison of deposited mass with calculated mass using the Haamaker equation.

The quality of electrophoretically deposited thick films was significantly improved in the presence of citric or oxalic acids. This can be seen in **Fig. 8.** While in the presence of oxalic or citric acids continuous thick films were deposited, cathodic depositions with nitric acid for zeta potential adjustment yielded non-homogeneously deposited material and sodium hydroxide yielded very sparse anodic depositions exhibiting a low density of adhered particles.

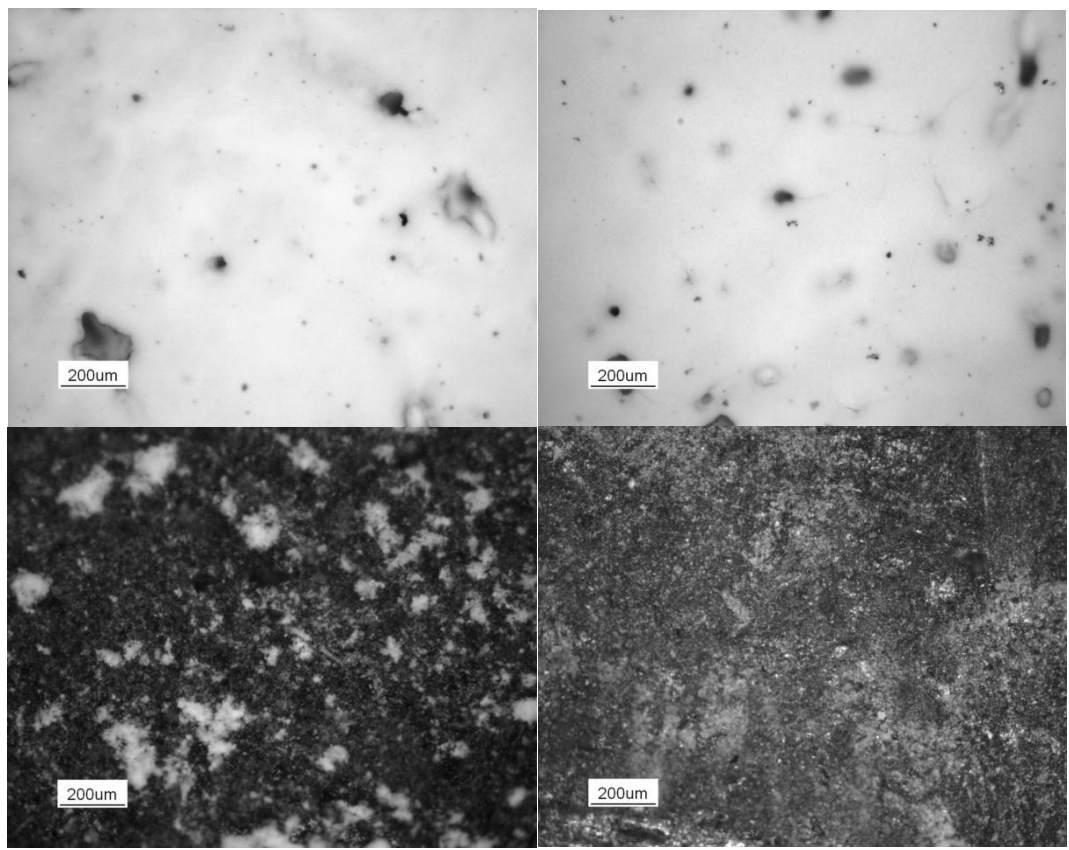

**Fig. 8.** Electrophoretically deposited $ZrO_2$ on graphite substrates from aqueous suspensions with (a) citric acid (b) oxalic acid (c) nitric acid and (d) sodium hydroxide additions.

## 4. Discussion

The data for conventional zeta potential variation with pH, as shown in Fig. 2, is similar to that of other systems and consistent with data in the literature. [15, 26] In contrast, as shown in Fig. 3, when citric acid was used for acidity adjustment, the zeta potential/pH curve shifted downwards by ~40 mV (relative to the values in Fig.2), showing only negative values for the zeta potential at pH levels higher than 2, similar to that of polyelectrolyte dispersants. [26, 42] This effect is attributed to the ligation of negative carboxylate groups to particle surfaces, in similarity to the case for alumina particles in aqueous suspensions. [11, 22] Further, while the use of conventional pH variation results in the expected observation of electrostatic dispersion at zeta potential values greater than ~30 mV and agglomeration at lower zeta potential values, the presence of citric acid maintained dispersion of $ZrO_2$ at all pH levels used in the present work. At low pH levels achieved with the use of citric acid, a zeta potential value of ~18 mV was observed without the occurrence of agglomeration. In the case of electrostatic dispersion such values of zeta potential are typically not sufficient to maintain particle dispersion, [37] and indeed suspensions of similar zeta potentials achieved with conventional pH adjustment (Fig. 2) were observed to show significant agglomeration. This is evidence that with the use of citric acid, in addition to the enhanced electrostatic repulsion imparted by a larger zeta-potential values, a steric barrier prevents particle approach and agglomeration, thus the mechanism of dispersion by citric acid is electrosteric rather than simply electrostatic. The stabilising effect of the steric barrier depends on the size of the adsorbed molecule, and thus smaller carboxylate groups, such as oxalate are likely to impart smaller steric barriers. [42, 43]

Since the adsorption of carboxylate groups on the particle surfaces is the mechanism by which particle dispersion was achieved, this effect was studied with the use of controlled concentrations of carboxylic acids of variable sizes, number of available carboxyl (anchor) groups, and speciation characteristics. Model structures of the three carboxylic acids, exhibiting symmetrical stereochemistry, are shown in **Fig. 9.**

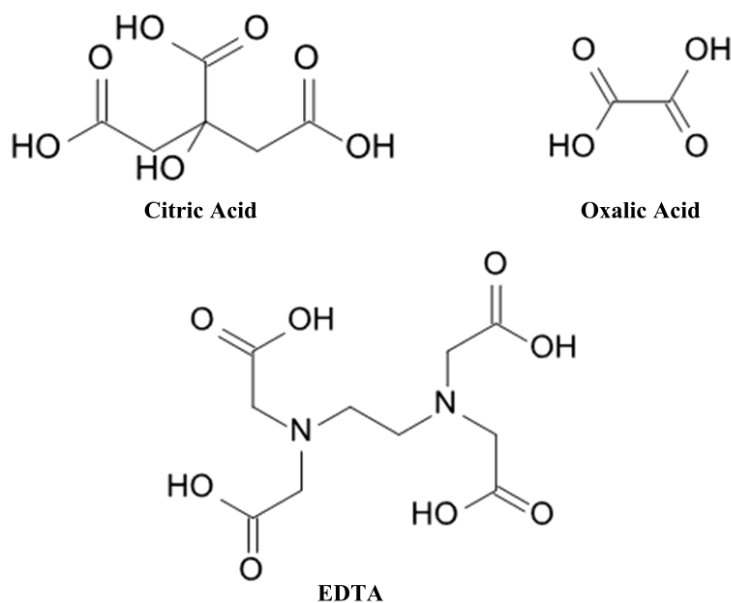

**Fig. 9. Molecular structures of carboxylic acids used in the present work**

As shown in Table 2, reported p$K_a$ values were used to determine the expected speciation behaviour at the pH levels used in the present work using Hyperquad Simulation and Speciation (HySS) software.[44] Speciation diagrams are shown in **Fig. 10.**

Table 2. Aqueous dissociation characteristics of carboxylic acids used in the present work [45-47]

| Reagent | p$K_{a1}$ | p$K_{a2}$ | p$K_{a3}$ | Expected Speciation at pH=4 | Expected Speciation at pH=5.4 |
|---|---|---|---|---|---|
| Citric Acid | 3.14 | 4.77 | 6.40 | 76% $AH_2^-$<br>13% $AH^{2-}$<br>11% $AH_3$ | 75% $AH^{2-}$<br>17% $AH_2^-$<br>8% $A^{3-}$ |
| Oxalic Acid | 1.25 | 4.14 | N/A | 57% $AH^-$<br>43% $A^{2-}$ | 94% $A^{2-}$<br>6% $AH^-$ |
| EDTA | 2.00 | 2.67 | 6.16 | 95% $AH_2^{2-}$<br>4% $AH_3^-$<br>1% $AH^{3-}$ | 85% $AH_2^{2-}$<br>15% $AH^{3-}$ |

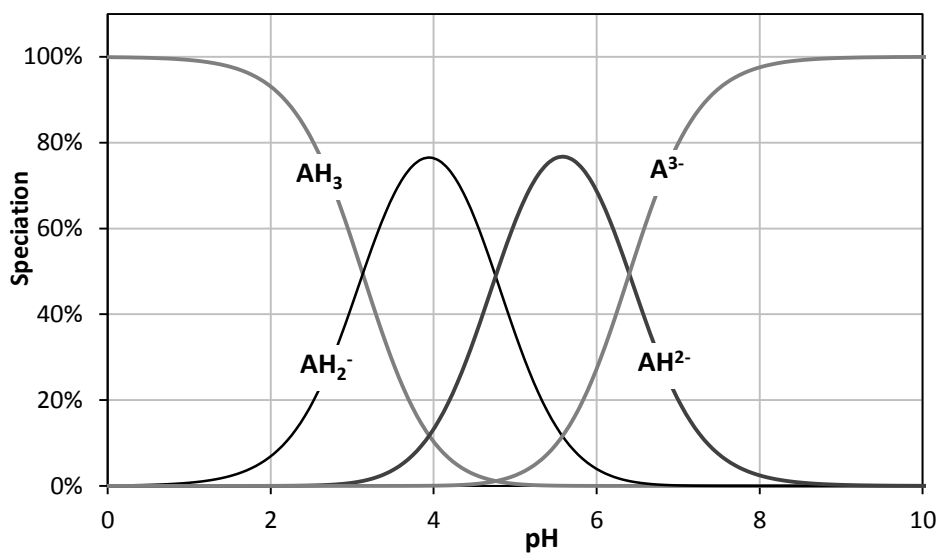

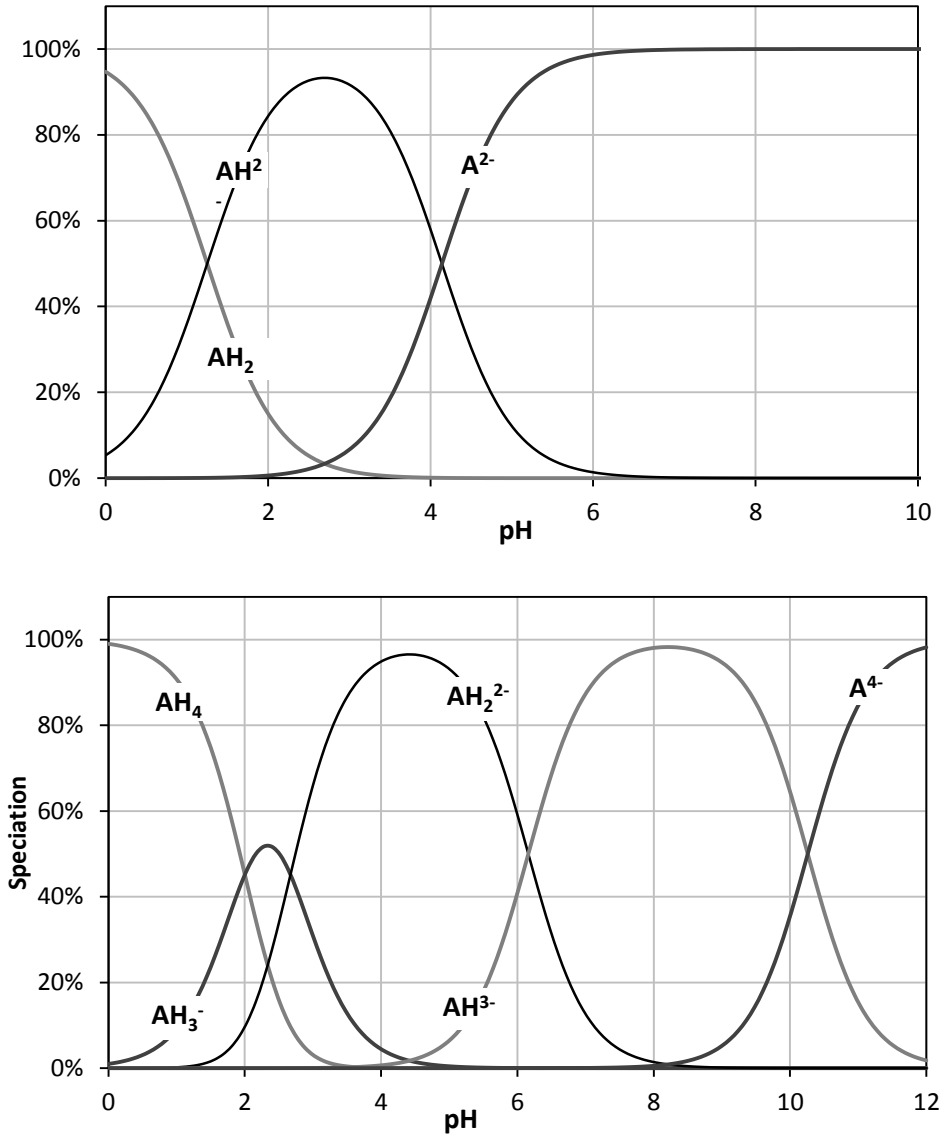

**Fig. 10. Aqueous speciation diagrams of (a) Citric acid (b) Oxalic acid and (c) EDTA**

Fig. 4 shows the effect of carboxylic acid additions on the zeta potential of a suspension at pH ~4 and Fig. 5 shows the effect of the same reagents on a $ZrO_2$ suspension at pH ~5.4, the native pH of the suspension. The decrease in zeta potential of 50-55 mV supports the conclusion of the ligation of the negative carboxylate groups and the levelling of the zeta potential drop corresponds to the surface saturation by these groups. The zeta potential and additive concentration at which the latter occurs depend principally on the surface charge imparted by the adsorbed species and their adsorption cross section on the particle surface. The effects of the carboxylic acids on zeta potential can thus be discussed in terms of size, surface coordination and speciation of the reagents used.

With the addition of citric acid and EDTA particle surface saturation is achieved at similar additive concentrations while with oxalic acid this occurs at higher addition levels. It is probable that a greater amount of oxalic acid was adsorbed owing to its smaller effective size and thus a smaller adsorption cross section relative to the other two reagents. Despite the larger molecular size of EDTA relative to citric acid, the adsorption cross section may be similar due to a similar surface complexation with $ZrO_2$ surfaces. Considering the surface area of the $ZrO_2$ powder, as determined from $N_2$ adsorption isotherms using BET methods, it is apparent that the surfaces monoclinic $ZrO_2$ reach saturation with citric acid / EDTA at ~1.1 µMol m$^{-2}$ while oxalic acid saturates particle surfaces at ~2.4 µMol m$^{-2}$. This corresponds to an adsorption cross section area of ~1.5 nm$^2$ for citric and EDTA and~ 0.7 nm$^2$ for oxalic acid. This value is somewhat larger than the adsorption cross section reported for citric acid on alumina surfaces. [22]

The adsorption of citric acid to oxide particle surfaces is reported to take place through ligand exchange of surface-adsorbed hydroxyl groups with two carboxylate groups. [22, 36, 48] As evident from its adsorption cross section, similar surface complexation is likely to take place with EDTA. From consideration of the pK$_a$ values and consequent speciation characteristics of EDTA and citric acid as shown in Fig. 10, it would be expected that, contrary to observed behaviour, EDTA would impart a stronger negative charge to $ZrO_2$ particles. However, as reported elsewhere, [36] adsorbed organic acids exhibit lower acidity constants and greater deprotonation relative to species in solution and thus the negative charges imparted by $ZrO_2$ adsorbed carboxylate groups are greater in magnitude than those predicted by Fig. 10. Taking into account the presence of the hydroxyl group in citric acid, greater negative charge density on particle surfaces may be achievable with this reagent. From considerations of structure and speciation, it is likely that fully deprotonated oxalate groups coordinate to particle surfaces with one carboxylate group, giving rise to a lower adsorption cross section and similar surface charge density to citric acid.

An increase in pH occurs as citric acid adsorbs to particle surfaces reaching a maximum as particle surfaces reach saturation, as shown in Fig. 6. This is most likely due to the displacement of hydroxyl groups from $ZrO_2$ surfaces. Further, it is possible that due to increased negative surface charges, more positively charged species in solution become trapped in a thicker electrical double layer surrounding the particles thus leading to a higher pH reading.

Using conventional reagents (nitric acid and sodium hydroxide) to modify the zeta potential of particles in suspension resulted in poor electrophoretic deposition behaviour as shown in Figs 7 and 8. This occurred mainly as a result of significant settling of particles in suspension, which is an expected consequence of the large particle size of the powder used in this work shown in Fig. 1. The electrophoretic deposition of thick $ZrO_2$ films from the powder used in this work was made possible when citric acid and oxalic acids were used as dispersants. This is further evidence that in addition to the modification of the zeta potential, carboxylic acids facilitate enhanced dispersion through steric barriers.

## Conclusions

Carboxylic acids have been shown to act as effective low molecular weight dispersants for aqueous suspensions of $ZrO_2$. The use of a small quantity of a carboxylic acid reagent is sufficient to cover particle surfaces with negatively charged species, impart highly negative zeta potential values and maintain the oxide particles in suspension over a wide range of pH levels without the onset of agglomeration. The formation of a steric barrier is likely to also be a contributing factor in maintaining stabilization of particle suspensions. The use of carboxylic dispersing reagents has been shown to highly be beneficial in the electrophoretic deposition of $ZrO_2$ thick films, enabling the deposition of micron-range particles.

## Acknowledgements

The authors wish to acknowledge assistance from the Institute of Materials Engineering at the Australian Nuclear Science and Technology Organisation (ANSTO) and the support of the Australian Institute of Nuclear Science and Engineering (AINSE).